\documentclass[aps,showpacs,unsortedaddress,twocolumn]{revtex4}

\usepackage{epsfig,amssymb,amsfonts,verbatim,enumerate}

\begin{document}
\font\Bbb =msbm10  scaled \magstephalf \def\id{{\hbox{\Bbb I}}}
\newcommand{\ket}[1]{| #1 \rangle}
\newcommand{\bra}[1]{ \langle #1|}
\newcommand{\proj}[1]{\ket{#1}\bra{#1}}
\newcommand{\braket}[2]{\langle #1|#2\rangle}
\newcommand{\half}{\mbox{$\textstyle \frac{1}{2}$}}
\def\opone{\leavevmode\hbox{\small1\kern-3.8pt\normalsize1}}
\newcommand{\tr}[1]{\mbox{Tr} \, #1 }
\def\emph#1{{\it #1}}
\def\textbf#1{{\bf #1}}
\def\textrm#1{{\rm #1}}

\title{Correlated Photon Emission from a Single II-VI Quantum Dot}

\author{C. Couteau}

\affiliation{CEA-CNRS-UJF Joint Group Nanophysics and
Semiconductors, Laboratoire de Spectrom\'{e}trie Physique, CNRS
UMR 5588, \\ Universit\'{e} J. Fourier Grenoble 1, 38402
Saint-Martin d'H\`{e}res, France}

\author{S. Moehl}\author{F. Tinjod}
\author{J.M. G\'{e}rard} \author{K. Kheng}\author{H. Mariette}

\affiliation{CEA-CNRS-UJF Joint Group Nanophysics and
Semiconductors, CEA/DRFMC/SP2M, 17 rue des Martyrs, 38054
Grenoble, France}

\author{J. A. Gaj}

\affiliation{Institute of Experimental Physics, Warsaw University,
Hoÿza 69, 00-681 Warsaw, Poland.}

\author{R. Romestain}\author{J.P. Poizat}

\affiliation{CEA-CNRS-UJF Joint Group Nanophysics and
Semiconductors, Laboratoire de Spectrom\'{e}trie Physique, CNRS
UMR 5588, \\ Universit\'{e} J. Fourier Grenoble 1, 38402
Saint-Martin d'H\`{e}res, France}

\begin{abstract}
We report correlation and cross-correlation measurements of
photons emitted under continuous wave excitation by a single II-VI
quantum dot (QD) grown by molecular-beam epitaxy. A standard
technique of microphotoluminescence combined with an ultrafast
photon correlation set-up allowed us to see an antibunching effect
on photons emitted by excitons recombining in a single CdTe/ZnTe
QD, as well as cross-correlation within the biexciton
($X_{2}$)-exciton ($X$) radiative cascade from the same dot. Fast
microchannel plate photomultipliers and a time-correlated single
photon module gave us an overall temporal resolution of 140 ps
better than the typical exciton lifetime in II-VI QDs of about
250ps.
\end{abstract}

\maketitle

The generation of triggered single photons has brought a lot of
interest in the last few years owing to its application in quantum
cryptography \cite{beveratos02,waks02} and its possible use for
quantum computation \cite{knill01}. Recent experiments have shown
that self-assembled quantum dots (SAQDs) are potentially a good
candidate for the production of triggered single photons
\cite{waks02,gayral99,michler00,sebald02}. In order to implement
Knill et al.'s proposal \cite{knill01}, one of the very stringent
requirement is to produce indistinguishable single photons to
enable multi-photon interference effects \cite{natsantori02}. The
possibility of creating on demand a single polarization entangled
photon pair directly from a SAQD has also been proposed
\cite{benson00}. In this case, the production of the entangled
pair of photons comes from the recombination of two excitons
within the same QD \cite{moreau01}. Polarization correlation
measurements of the biexcitonic ($X_2$) cascade to the ground
state via the exciton ($X_1$) state have been reported recently in
III-V \cite{santori02} and II-VI \cite{ulrich03} compounds.
However, no entanglement has been seen so far in the $X_2$
cascade.

Most of the results on single photon sources using SAQDs have been
obtained with III-V semiconductors. Only a few recent papers
\cite{sebald02,aichele03,ulrich03} present photon correlation
measurements on II-VI SAQDs, all with CdSe/ZnSe. The family of
II-VI materials exhibits interesting features for quantum
information applications. The lifetime (of the order of $200$ ps)
is smaller than in III-V SAQDs. It is an advantage for the
production of indistinguishable single photons since this means
that the emission of a photon takes less time and should then be
less sensitive to dephasing. This short lifetime could also allow
higher repetition rate (in the GHz range).

In this letter, we report the observation of antibunching of the
excitonic line of CdTe/ZnTe SAQD, and the biexciton-exciton
cross-correlation under continuous wave (cw) excitation. The short
lifetimes in II-VI semiconductors have made it necessary to set-up
an ultrafast photon correlation experiment using rapid
microchannel plate photomultipliers.

The investigated sample consists of a CdTe SAQD layer (with dot
density of about $3~10^{10}cm^{-2}$) embedded in ZnTe barriers.
 A detailed description of the sample growth can be found in
\cite{tinjod03}. Single dot spectroscopy is obtained by exciting
the sample through submicron apertures (with sizes down to $0.2
{\mu}m$) made in a thin ($100$nm) aluminium mask deposited onto
the surface. The sample is mounted on the cold finger of a helium
flux cryostat keeping the QDs at about 5K for all the experiments
described here. A microscope objective (numerical aperture of
$0.4$) and a high refractive index ($n=2.16$) hemispherical solid
immersion lens (SIL) \cite{moehl02} are used to collect the single
dot emission efficiently. The $488$nm ($2.541$eV) line of a cw
argon laser is sent through the same optics for the excitation.
Fine positioning of the laser onto the chosen aperture in the mask
is obtained with submicronic displacements of the microscope
objective by XYZ piezo drivers.

The collected QD photoluminescence (PL) is then sent on a 50/50
beamsplitter for correlation measurements \cite{hbt56}. In each
arm of the beamsplitter, the light is dispersed by a monochromator
(grating of 1200 grooves/mm, 30cm and 50cm focal length
respectively). Each monochromator has a switchable mirror inside
that can direct the luminescence either onto a CCD camera for the
measurement of the PL spectrum or through the exit slit towards
the single photon counter. In our case these detectors are
microchannel plate photomultipliers that have a very small transit
time jitter ($< 50$ps). The detectors send electrical pulses into
a time-correlated single photon module that contains all the
electronics necessary for the data processing. The overall
temporal resolution of our set-up is essentially limited by the
jitter of the detectors and the dispersion of the monochromator
gratings. This time resolution was measured by recording the
autocorrelation function of 200fs long pulses from a
frequency-doubled Ti:Sapphire laser at 80MHz repetition rate. A
FWHM of $140$ps was obtained for the autocorrelation function
peaks. This is, to our knowledge, the shortest time resolution
ever reported for photon correlation measurements.

\begin{figure}
\begin{center}
\includegraphics[scale=0.7]{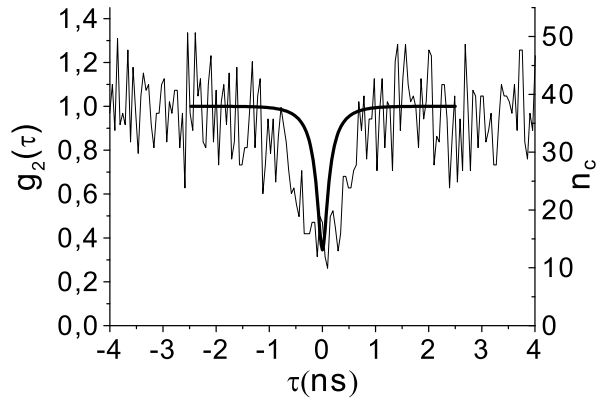}
\caption{\footnotesize{\textit{Autocorrelation measurement of a
single exciton. The left axis is the autocorrelation function
$g^{(2)}(\tau)$ and the right axis is the unormalized coincidence
rate. The time bin is 49ps and  $n_c$ is the total number of
events during the 12600s of acquisition. Typical counting rates
were $n_{start}=10~000$cps and $n_{stop}=7~000$cps on a single
exciton line. The background count rates of each channel were
$n_{start}^{d}=1~000$cps and $n_{stop}^{d}=800$cps. The boldline
is the expected curve for a multiexcitonic model, see text.}}}
\label{Fig:g2cdte}
\end{center}
\end{figure}

Fig \ref{Fig:g2cdte} shows the autocorrelation function
$g^{(2)}(\tau)$ of a QD excitonic line at $2.232$eV ($555.5$nm).
It displays the reduced coincidence rate around $\tau = 0$
(antibunching), which is characteristic of the photon statistics
of a single emitter. We have chosen to study excitonic lines
 on the high energy side of
the CdTe QDs emission distribution because the sensitivity
 of our detectors drops very quickly at
lower energies. In these measurements, we have performed a careful
calibration of the pump power in unit of $\Gamma_{X_1}=1/T_{X_1}$
where $T_{X_1}$ is the lifetime of the exciton. This has been done
by fitting the luminescence intensity versus the pump power using
a multiexciton ladder model \cite{regelman01} with a population of
up to 4 excitons in the dot. In this model the lifetimes of the
first two excitons ($X_1$ and $X_2$) were taken from experimental
measurements (see below), and the lifetimes of the triexciton
($X_3$) and the quadriexciton ($X_4$), which are not very
critical, were taken from ref \cite{besombes02}. The
autocorrelation function of Fig \ref{Fig:g2cdte} was obtained with
a QD excitation rate $r=0.65\Gamma_{X_1}$.

After convolution with the time resolution of our set-up, and
taking into account the signal to background ratio
\cite{Brouri00}, the bold line in Fig \ref{Fig:g2cdte} is the
prediction of the ladder model for the autocorrelation function.
There is a difference of a factor of 4 between the FWHM of the
experimental curve and the model prediction. This mismatch has
been observed on the three quantum dots of the same sample that we
have studied. Previous groups \cite{becher01,moreau02} have
observed the same mismatch in the past for III-V compounds. The
presence of dark excitons may explain this discrepancy as
suggested by the measurements described in the following.

\begin{figure}
\begin{center}
\includegraphics[scale=0.7]{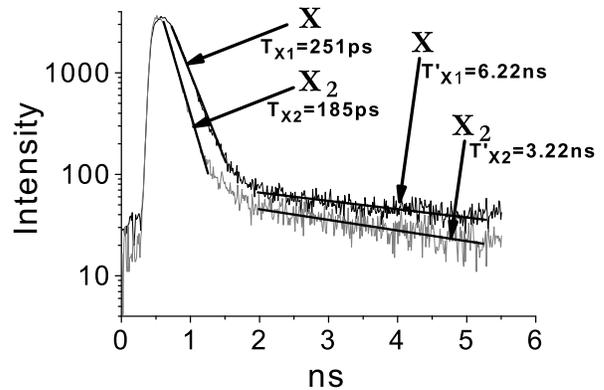}
\caption{\footnotesize{\textit{Lifetime measurements of the
exciton and its associated biexciton done with a pulsed
Ti:Sapphire laser.}}} \label{Fig:decaycdte}
\end{center}
\end{figure}

Decay time measurements (see Fig \ref{Fig:decaycdte}) using a
pulsed Ti:Sapphire laser (pulse duration 1ps) give lifetime values
of $T_{X_1}=251\pm5$ps and $T_{X_2}=185\pm4$ps for the exciton and
its associated biexciton respectively. A biexponential decay is
observed both for $X_1$ and $X_2$, with $T'_{X_1}=6.22\pm{0.68}$ns
and $T'_{X_2}=3.22\pm{0.16}$ns for their slow components. This
longer time decay might be the signature of the presence of the
dark exciton as seen in \cite{labeau03}. Nevertheless, further
investigations, like temperature dependent measurements, need to
be done to resolve the question.

\begin{figure}
\begin{center}
\includegraphics[scale=0.7]{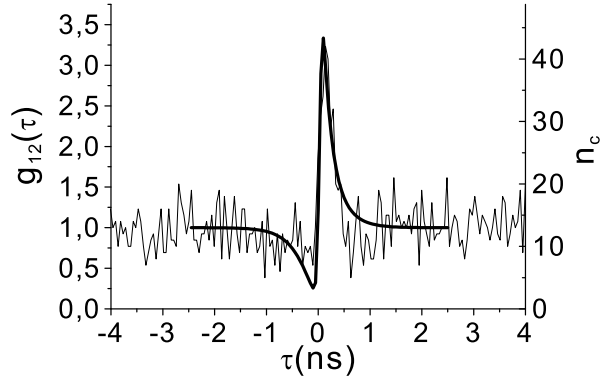}
\caption{\footnotesize{\textit{Cross-correlation measurement
between an exciton and its biexciton.The bold line is the expected
curve for the same multiexcitonic model, see text for details.
$n_c$ is again the total number of events during the 9~720s of
acquisition. The time bin is 49ps. Counting rates are
$n_{2}=4~800$cps for the $X_{2}$ and $n_{1}=7~600$cps for the
$X_{1}$ with background count rates of $n^{d}_{2}=400$cps and
$n^{d}_{2}=600$cps .}}} \label{Fig:g12cdte}
\end{center}
\end{figure}

Fig \ref{Fig:g12cdte} shows the correlation measurement between
the same exciton and its biexciton \cite{regelman01,moreau01}. For
that, we used the same set-up, but one of the spectrometers was
tuned on the wavelength of the biexcitonic line at 2.219eV
(558.6nm), that is 13 meV below the single exciton emission.
 The pump parameter was
$r=0.204\Gamma_{X_1}$.  When $X_{2}$ triggers the \textit{start}
and $X_{1}$ stops the time counter, a discontinuity at $\tau=0$ is
expected with value above unity for $\tau>0$ revealing the
bunching part of the cascade \cite{moreau01}.  For $\tau<0$,
antibunching is expected, since a recycling time is necessary for
a biexciton photon to be emitted after the last exciton has been
recombined.

Our result in Fig \ref{Fig:g12cdte} displays very clearly the
bunching part for positive times, and less clearly the
antibunching part for negative times. The bold line is the fitting
curve using the same ladder model with up to 4 excitons in the
dot. Now, we observe that, contrary to the antibunching
measurement in Fig \ref{Fig:g2cdte}, we do have the correct decay
time corresponding to $T_{X_1}$ for  positive times. The
discrepancy of a factor of 4 for the exciton autocorrelation is no
longer present in cross-correlation measurements. This indeed
suggests an influence of dark excitons which would explain the
discrepancy of Fig \ref{Fig:g2cdte}, since in cross-correlation
experiments the recombination of a biexciton always leaves an
exciton in a bright state.

To conclude, thanks to our temporal resolution, we presented in
this letter  cw autocorrelation and cross-correlation measurements
of a single self-assembled quantum dot made of II-VI materials. We
reported expected and unexpected behaviors from these results as
discussed above. Further experiments using pulsed excitation need
doing in CdTe/ZnTe QDs but the first results shown here present a
good omen that such heterostructures might be good candidates for
the purposes of quantum optics and quantum information.

We thank Ph. Grangier for the loan of the two microchannel plate
photomultipliers,  M. Terrier for the  processing of the sample,
and F. Donatini for experimental assistance. This work is
supported by the French Ministry for Research (ACI "Polqua", and
ACI Jeune Chercheur) and the "Sciences et Technologies de
l'Information et de la Communication" Department of the CNRS.
J.A.G. acknowledges the support of Polish Committee for Scientific
Research (grant No PBZ-KBN-044/P03/2001).

We would like to dedicate this paper to the memory of Robert
Romestain who died in a mountain accident during the writting of
this paper.


\begin{thebibliography}{99}

\bibitem{beveratos02} A. Beveratos, R. Brouri, T. Gacoin,
A. Villing, J.Ph. Poizat, and Ph. Grangier, Phys. Rev. Lett. {\bf
89}, 187901 (2002)

\bibitem{waks02} E. Waks, K. Inoue, C. Santori, D. Fattal, J. Vuckovic, G.S.
Solomon, and Y. Yamamoto,Nature (London) {\bf 420}, 762 (2002).

\bibitem{knill01} E. Knill, R. Laflamme, and G.J. Milburn, Nature (London) {\bf 409}, 46 (2001).

\bibitem{gayral99} J.M. G\'erard and B. Gayral, J. Lightwave
Technol. \textbf{17}, 2089 (1999).

\bibitem{michler00} P. Michler, A. Kiraz, C. Becher, W.V. Schoenfeld,
P.M. Petroff, L. Zhang, E. Hu, and A. Imamoglu, Science {\bf290},
2282 (2000).

\bibitem{sebald02} K. Sebald, P. Michler, T. Passow, D. Hommel, G.
Bacher, and A. Forchel, Apll. Phys. Lett. \textbf{81}, 2920
(2002).

\bibitem{natsantori02} C. Santori, D. Fattal, J. Vuckovic, G. S. Solomon, and Y. Yamamoto,
Nature (London) {\bf419}, 594 (2002).


%\bibitem{santori01} C. Santori, M. Pelton, G. S. Solomon, Y. Dale,
%and Y. Yamamoto, Phys. Rev. Lett. {\bf86}, 1502 (2001).

\bibitem{benson00} O. Benson, C. Santori, M. Pelton, and Y. Yamamoto,
Phys. Rev. Lett. {\bf84}, 2513 (2000).



% \bibitem{moreauapl01} E. Moreau, I. Robert, J. M. Gérard, I. Abram,
% L. Manin, and V. Thierry-Mieg,
% Appl. Phys. Lett. \textbf{79}, 2865 (2001)
%Single-mode solid-state single photon source based on isolated
% quantum dots in pillar microcavities

\bibitem{moreau01} E. Moreau, I. Robert, L. Manin, V. Thierry-Mieg,
J.M. G\'{e}rard, and I. Abram, Phys. Rev. Lett. {\bf87}, 183601
(2001).

\bibitem{santori02} C. Santori, D. Fattal, M. Pelton, G. S. Solomon, and Y. Yamamoto,
Phys. Rev. B {\bf66}, 45308 (2002).


\bibitem{ulrich03} S.M. Ulrich, S. Strauf, P. Michler, G. Bacher, and A. Forchel,
App. Phys. Lett. {\bf83}, 1848 (2003).


\bibitem{aichele03} T. Aichele, V. Zwiller, O. Benson, I. Akimov,
and F. Henneberger, J. Opt. Soc. Am B \textbf{20}, 2189 (2003).


%\bibitem{gerard98} J.M. G\'{e}rard, B. Sermage, B. Gayral, B. Legrand, E. Costard,
%and V. Thierry-Mieg, Phys. Rev. Lett. {\bf 81}, 1110 (1998).

%\bibitem{besombes01} L. Besombes, K. Kheng, L. Marsal, and H.
%Mariette, Phys. Rev. B {\bf 63}, 155307 (2001).

%\bibitem{favero03} I. Favero, G. Cassabois, R. Ferreira, D. Darson,
%C. Voisin, J. Tignon, C. Delalande, and G. Bastard, Phys. Rev. B
%{\bf 68}, 233301 (2003).

\bibitem{tinjod03} F. Tinjod, B. Gilles, S. Moehl, K. Kheng, and H. Mariette,
App. Phys. Lett. {\bf82}, 4340 (2003).


%\bibitem{glauber63} R. J. Glauber, Phys. Rev. {\bf130}, 2529 (1963).

%\bibitem{beveratos02} A. Beveratos, S. K\"uhn, R. Brouri, T. Gacoin,
%J.P. Poizat, and P. Grangier,
%Eur. Phys. J. D {\bf18}, 191 (2002).

\bibitem{moehl02} S. Moehl, Hui Zhao, B. Dal Don, S. Wachter, and H.
Kalt, J. App. Lett. {\bf93}, 6265 (2002); W.L. Barnes, G. Bj\"ork,
J.M. G\'erard, P. Jonsson, J.A.E Wasey, P.T. Worthing, and V.
Zwiller, Eur. Phys. J. D \textbf{18}, 197 (2002).

\bibitem{hbt56}R. Hanbury Brown, and R. Q. Twiss,
Nature (London) {\bf177}, 27, (1956).


%\bibitem{dekel00} E. Dekel, D. Gershoni, E. Ehrenfreund, J.M. Garcia, and P.M. Petroff, Phys. Rev. B
%{\bf61}, 11009 (2000).

\bibitem{regelman01} D.V. Regelman, U. Mizrahi, D. Gershoni, E.
Ehrenfreund, W.V. Schoenfeld, and P.M. Petroff, Phys. Rev. Lett.
{\bf87}, 257401 (2001).

\bibitem{besombes02} L. Besombes, Th\`{e}se de Doctorat, Universit\'{e}
Joseph Fourier Grenoble (2002).

\bibitem{Brouri00} R. Brouri, A. Beveratos, J.Ph. Poizat, and Ph.
Grangier, Opt. Lett. \textbf{25}, 1294 (2000).

\bibitem{becher01} C. Becher, A. Kiraz, P. Michler, A. Imamoglu, W.V. Schoenfeld,
P.M. Petroff, L. Zhang, and E. Hu, Phys. Rev. B {\bf 63}, 121312
(2001).

\bibitem{moreau02} E. Moreau, Th\`{e}se de Doctorat, Universit\'{e}
Paris-Sud Orsay (2002).

\bibitem{labeau03} O. Labeau, P. Tamarat, and B. Lounis, Phys. Rev. Lett.
{\bf90}, 257404 (2003).

\end{thebibliography}
\end{document}